\documentclass[apj, iop]{emulateapj}
\usepackage{apjfonts}
\usepackage{natbib}
\usepackage{ulem}
\usepackage[dvipdfm,breaklinks,colorlinks,citecolor=blue]{hyperref}

\newcommand{\MO}{$M_\odot$}

\newcommand{\Chandra}{\textit{Chandra}}
\newcommand{\XMM}{\textit{XMM-Newton}}
\newcommand{\NH}{$N_{\rm H}$}

\received{}
\revised{}
\accepted{}

\shorttitle{Embedded Spiral Patterns in the Cool Core of Abell 1835}
\shortauthors{Ueda et al.}

\begin{document}

\title{Embedded Spiral Patterns in the Cool Core of the Massive Cluster of Galaxies Abell\,1835}

\email{sueda@astro.isas.jaxa.jp}

\author{Shutaro Ueda}
\affiliation{Institute of Space and Astronautical Science (ISAS), Japan Aerospace Exploration Agency (JAXA) \\
3-1-1 Yoshinodai, Chuo \\
Sagamihara, Kanagawa 252-5210, Japan}

\author{Tetsu Kitayama}
\affiliation{Department of Physics, Toho University \\
2-2-1 Miyama \\
Funabashi, Chiba 274-8510, Japan}

\author{Tadayasu Dotani}
\affiliation{Institute of Space and Astronautical Science (ISAS), Japan Aerospace Exploration Agency (JAXA) \\
3-1-1 Yoshinodai, Chuo \\
Sagamihara, Kanagawa 252-5210, Japan}
\affiliation{SOKENDAI (The Graduate University for Advanced Studies) \\
3-1-1 Yoshinodai, Chuo \\
Sagamihara, Kanagawa 252-5210, Japan}






\begin{abstract}
We present the properties of intracluster medium (ICM) in the cool core of the massive cluster of galaxies Abell\,1835 obtained with the data by {\it Chandra X-ray Observatory}. 
We find distinctive spiral patterns with the radius of 70\,kpc (or 18\arcsec) as a whole
in the residual image of X-ray surface brightness\ after the 2-dimensional ellipse model of surface brightness is subtracted.
The size is smaller by a factor of 2 -- 4  than that of other clusters known to have a similar pattern.
The spiral patterns consist of two arms.  
One of them appears as positive, and the other does as negative excesses in the residual image.
Their X-ray spectra show that the ICM temperatures in the positive- and negative-excess regions are $5.09^{+0.12}_{-0.13}$\,keV and  $6.52^{+0.18}_{-0.15}$\,keV, respectively.  
In contrast, no significant difference is found in the abundance or pressure, the latter of which suggests that the ICM in the two regions of the spiral patterns is  in pressure equilibrium or close.
The spatially-resolved X-ray spectroscopy of the central region ($r<40\arcsec$) divided into 92 sub-regions indicates that Abell\,1835 is a typical cool core cluster.
We also find that the spiral patterns extend from the cool core out to the hotter surrounding ICM.
The residual image reveals some lumpy sub-structure in the cool core.  
The line-of-sight component of the disturbance velocity responsible for the sub-structures is estimated to be lower than 600\,km\,s$^{-1}$.
Abell 1835 may be now experiencing an off-axis minor merger.
\end{abstract}





\section{Introduction}

Clusters of galaxies are the largest gravitationally-bounded objects in the Universe.
Cold dark matters form a gravitational potential well of a cluster.
The hot and optically thin plasma called intracluster medium (ICM) fills at the typical temperature of $10^{7} - 10^{8}$\,K in the well.
At the bottom of the well, a giant elliptical galaxy, which is often called the brightest cluster galaxy (BCG), is situated in most of clusters.
Since the density of the ICM in the vicinity of the BCG is much higher than that in the outskirts, its cooling time is shorter than the lifetime of clusters.
Then, a cool core is formed and it was speculated that the cooled ICM would induce massive starburst in the vicinity of BCG \citep{Fabian94}.
However, a massive starburst has not been observed in most of clusters to date except for few \citep[e.g.,][]{McDonald12}.

Clusters of galaxies are also a dynamically young system in the Universe under the scheme of hierarchical evolution.
They have evolved through mergers between smaller clusters.
Mergers often generate irregular structures in the X-ray surface brightness of clusters, such as a sub-peak \citep[e.g.,][]{Arnaud00, Sun02}, a cold front \citep[e.g.,][]{Markevitch00, Vikhlinin01}, and an elongated profile \citep[e.g.,][]{Briel92, Allen02} \citep[see][for a review]{Markevitch07}.
On the other hand, relaxed clusters, which has a symmetrical X-ray surface brightness in its core, are thought not to have experienced a violent (major) merger in the last several Gyrs \citep[e.g.,][]{Poole06}.

Recent high angular-resolution observations with \Chandra ~and \XMM ~have discovered embedded spiral patterns with a size of a few hundreds kpc in the residual image of X-ray surface brightness of several clusters and groups of galaxies after their nominal profile is subtracted,
e.g., the Perseus cluster \citep{Churazov03}, Abell\,2052 \citep{Blanton11}, Abell\,2029 \citep{Clarke04, Paterno-Mahler13}, Abell\,496 \citep{Ghizzardi14}, Abell\,3581 \citep{Canning13}, PKS\,0745-191 \citep{Sanders14}, Abell\,85 \citep{Ichinohe15}, IRAS\,09104$+$4109 \citep{OSullivan12}, and NGC\,5044 \citep{OSullivan14}.
Those spiral patterns usually consist of two arc-like parts.
One of them comes out as positive excess in the residual image and the other as negative.
A brighter area in an observed X-ray surface brightness is usually interpreted as a region with a higher density of ICM than that in darker parts, and so probably is the positive-excess region in the residual X-ray image of a cluster.
Moreover, the spiral arms show a lower temperature and entropy ($\displaystyle K = kT \times n_{\rm e}^{-2/3}$) in the positive-excess region than in the negative-excess region.
The metal abundance in the positive-excess region is higher than that in the surroundings \citep{Blanton11, Paterno-Mahler13, Ghizzardi14}.
However, the spatial variation of pressure ($P = kT \times n_{e}$) is still unclear.
Numerical simulations suggest that an off-axis minor merger can generates such spiral patterns \citep[e.g.,][]{Ascasibar06, ZuHone10, ZuHone11, Roediger11}.
A possible scenario of its formation is as follows.
An infalling subcluster sloshes the gravitational potential well of the main cluster.
Then, the cool, dense, and metal-rich ICM at the center of the main cluster is lifted up to the outskirts and the hot ICM at the outskirts flows to the center, which creates the features commonly observed.
Since the inflow ICM has a relatively high entropy, the mixing transfers heat to the cooler ICM.
On the other hand, it is considered that the gas sloshing is one of the heating sources to suppress  cooling of the ICM \citep{Ghizzardi10, ZuHone10} in addition to the feedback of active galactic nucleus (AGN) in the BCG \citep{McNamara07}.

Although it is considered that embedded spiral patterns are closely related to the cool core, only a limited amount of studies have been made so far.
In particular, none of the published data are good enough to investigate the time evolution of spiral patterns.
The number of the sample of massive clusters is also small.
Mergers are usually associated with an evolution of the mass of the clusters.
We need more samples of distant and massive clusters to fully understand the evolution of spirals.

In this paper, we present our result of, in quest for possible evidence of past mergers, the cool core of a massive and relaxed clusters of galaxies Abell\,1835 (hereafter A1835) at the redshift of $z = 0.2532$ with the data of deep \Chandra ~observations.
A1835 is the most luminous X-ray cluster in the {\it ROSAT} Bright Cluster Sample \citep[BCS;][]{Ebeling98}, and thus is one of the most massive clusters.
It is classified as a relaxed and massive cooling-flow cluster.
Its X-ray surface brightness shows a strong central concentration \citep{Allen96}.
A cool core is formed in its center.
On the other hand, 
\cite{Bonamente13} and \cite{Ichikawa13} suggested that the ICM in the outskirts of A1835 is mostly out of hydrostatic equilibrium.
The global pattern of properties of the ICM is studied by \cite{Kirkpatrick15} and \cite{Hofmann16}
The BCG of A1835 has an exceptionally high star-formation rate of 100 -- 180\,\MO \,yr$^{-1}$ \citep{McNamara06} and thus is a rare object like the Phoenix cluster \citep{McDonald12}.
However, most of the ICM in the center is still hot \citep{Peterson01}.
The AGN in the BCG is not luminous in X-rays at all \citep{Russell13}, whereas moderately bright at radio wavelengths \citep{Korngut11}.
Although the signal of Sunyaev-Zel'dovich effect \citep[][for a review]{Sunyaev72, Sunyaev80, Kitayama14} was observed and detected by the MUSTANG \citep{Korngut11}, the signal in the central  region ($r < 18\arcsec$)  was strongly contaminated by the AGN emission.
\cite{Govoni09} suggested that a diffuse radio emission (i.e., a radio mini-halo) resides with a size of several hundreds kpc.
The total mass is estimated to be $M_{200} = 1.09 \times 10^{15} h^{-1}$\,\MO ~from a weak-lensing observation \citep{Okabe10}.
The mass reconstruction image from the weak-lensing shear signal shows a standard centrally-concentrated profile and no features of violent disturbance at least within $r < 10\arcmin$ \citep{Clowe02}.
These results  imply that A1835 shows no significant evidence of major merger.
Based on the observations of the mass fraction of the sub-structures, \cite{Smith08} suggested that this cluster was formed at $z \ge 0.8$ and has grown in mass by $\le 10$\,\% in the last 2\,Gyr at the cluster frame, which corresponds to $z \sim 0.4$.
Their results also confirmed that A1835 is a relaxed cluster.

Throughout the paper, we adopt the abundance table of \cite{Anders89}, the Hubble constant of $H_{0} = 70$\,km\,s$^{-1}$\,Mpc$^{-1}$, 
and the cosmological density parameters of $\Omega_{\rm M} = 0.27$ and $\Omega_{\Lambda} = 0.73$. 
Then, 1 arcsec corresponds to 3.97\,kpc at the redshift $z=0.2532$. 
All the errors quoted in this paper  are in 90\,\% confidence level (90\,\% CL) unless otherwise specified.

\section{Observations and Data Reductions}

\Chandra ~observations of A1835 with Advanced CCD Imaging Spectrometer \citep[ACIS;][]{Garmire03} were performed five times with the assigned ObsIDs of 495, 496, 6880, 6881, and 7370.
The observations of ObsID of 495 and 496 were carried out with the \Chandra/ACIS-S and the others were with the \Chandra/ACIS-I.
Among them, the data obtained with the ACIS-S were affected by a mild flare as reported by \cite{McNamara06}, and thus we do not use them in this paper.
Table~\ref{tab:logs}\ summarizes the observation log for the data used in this paper.
We reprocessed the event data using the Chandra Interactive Analysis of Observations \citep[CIAO;][]{Fruscione06} version 4.7 and the calibration database (CALDB) version 4.6.9.
The total net exposure is 193.7\,ksec.

\begin{table}[htbp]
\begin{center}
\caption{
Observation log of A1835 with \Chandra
}\label{tab:logs}
\begin{tabular}{rccr}
\hline\hline
ObsID		& Instrument 		& Obs. Date         & Expo. (ks) 	\\ \hline
6880		& ACIS-I				& 2005-12-07		& 117.9 	\\
6881		& ACIS-I				& 2006-08-25		& 36.3 	\\
7370		& ACIS-I 				& 2006-07-24 		& 39.5 	\\
\hline
\end{tabular}
\end{center}
\end{table}

\section{Analyses and Results}

\begin{figure*}[ht] 
\begin{center}
\includegraphics[width=60mm, height=60mm]{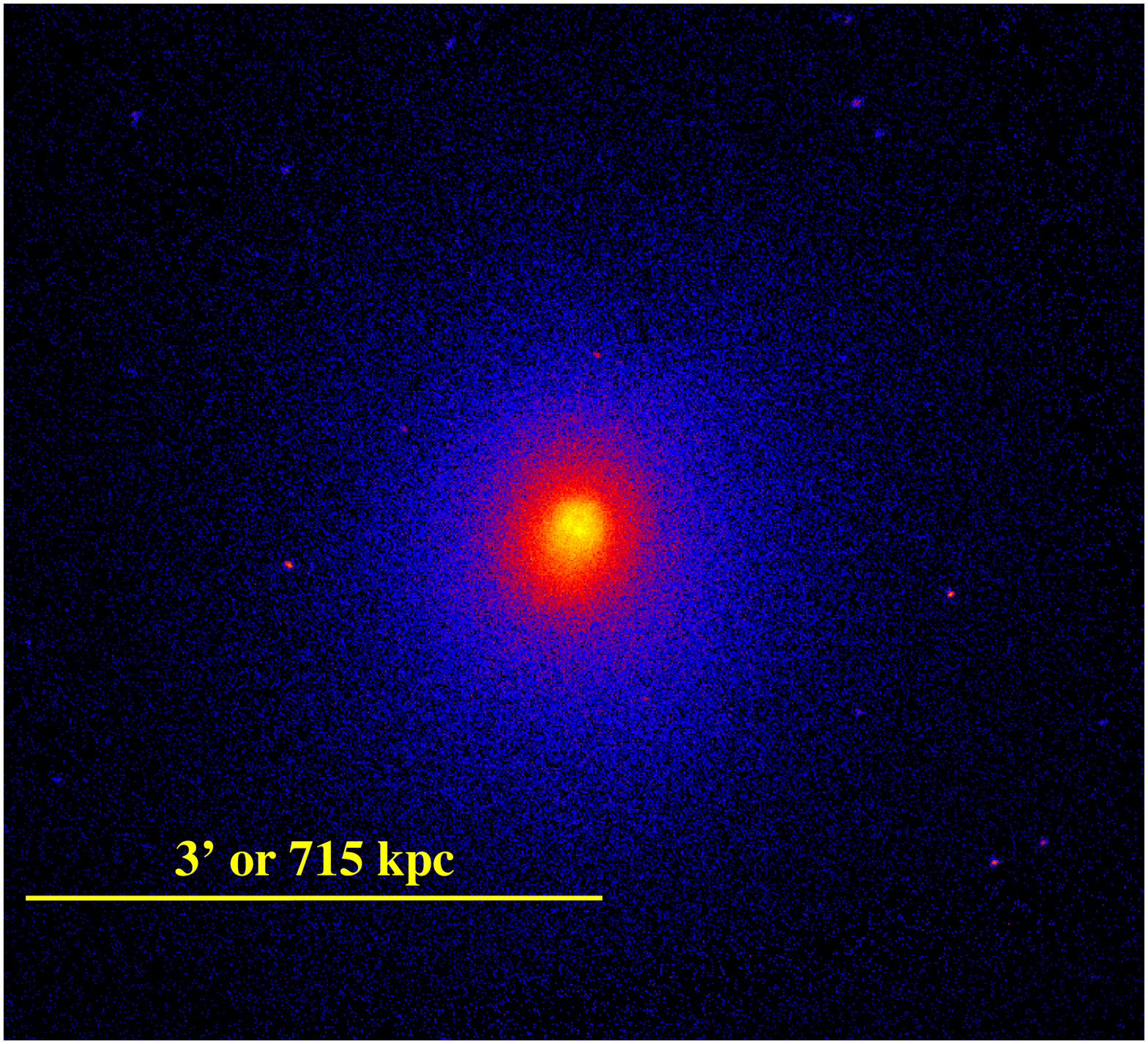}  
\hspace{5mm}
\includegraphics[width=70mm, height=70mm]{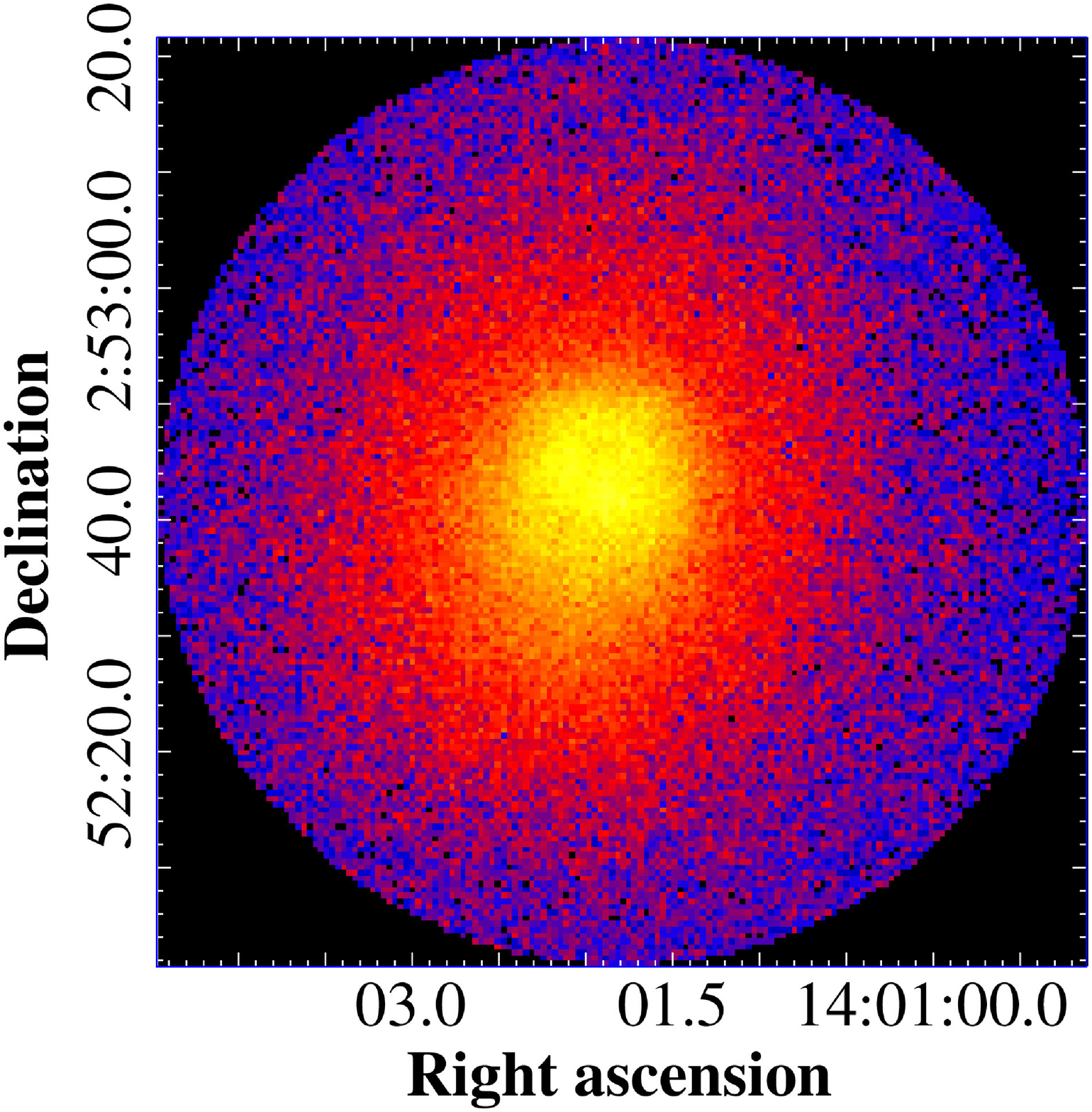}  
\caption{
X-ray image of A1835 in the energy range of 0.4 -- 7.0\,keV with the background subtracted and exposure corrected.
Left panel: large-scale image of A1835. The yellow bar indicates the scale of $3\arcmin$ or 715\,kpc.
This image shows a distinctive cool core in the center of A1835, which appears to be elliptical.
No evidence of major merger is found in this X-ray surface brightness.
Right panel: zoomed-up X-ray image of the central region of A1835 within $r < 40\arcsec$ or 159\,kpc from the left panel.
}
\label{fig:image}
\end{center}
\end{figure*}

We perform the X-ray imaging and spectral analyses to study the properties of the ICM in the central region of A1835.
We have chosen the annular region with the radii $r$ of $4\arcmin< r < 5\arcmin$ from the center of the cluster to extract the background data, which corresponds to over 1\,Mpc in distance from the center.
The cluster center is defined as the brightest pixel in the X-ray count map after all the data have been merged and is located at (RA, Dec) = (14h01m01.794, +02d52m41.55) in J2000.
We made the exposure-corrected X-ray image of A1835 in the energy range of $0.4 - 7.0$\,keV after subtracting the background as displayed in Figure~\ref{fig:image}.
No evidence of major merger is found in these images of X-ray surface brightness.
The X-ray profile appears to be elliptical and shows a cool core  in the cluster center, which suggests A1835 to be one of relaxed and cool-core clusters.

\subsection{Search and analyses of the spirals}
\label{sec:search}

Even if a cluster is classified as a relaxed and cool-core cluster, some subtle and possibly complex features may exist, which are often embedded under a global image profile of the cluster and can be easily overlooked. 
Detailed image analysis with a residual image of its core, after the global profile is subtracted from, may well highlight those features (see Introduction for some examples).
In this subsection, we search for such embedded features using a residual image.

First, we approximate the X-ray image profile of the cluster with the 2-dimensional ellipse model.
To determine the ratio of major-axis to minor-axis ($f$) and angle ($\theta$) of the ellipse model, 
we optimize $f$ and $\theta$ in the ranges of $0 < f \le 1$ and $0 \le \theta < 180$, respectively, 
to minimize the sum of RMS/mean, where  ``mean'' is the average photon counts in each isophote ring and  ``RMS''  is the root-mean-square difference between the model and the photon counts in the same ring.
As a result, we obtain $f = 0.857$ and $\theta = 170.9$, respectively.
Our definition is such that $\theta = 0$ and 90 mean the directions of south and west, respectively.
The mean surface brightness at each spot is thus given by the mean photon counts in the corresponding isophote ring in the ellipse model.
The derived ratio ($f$) is consistent with that derived by \cite{Schmidt01} with the 30\,ks exposure data of an early observation of {\it Chandra} and those of typical relaxed clusters reported by \cite{Kawahara10}.

\begin{figure*}[ht] 
\begin{center}
\includegraphics[width=64mm, height=58mm]{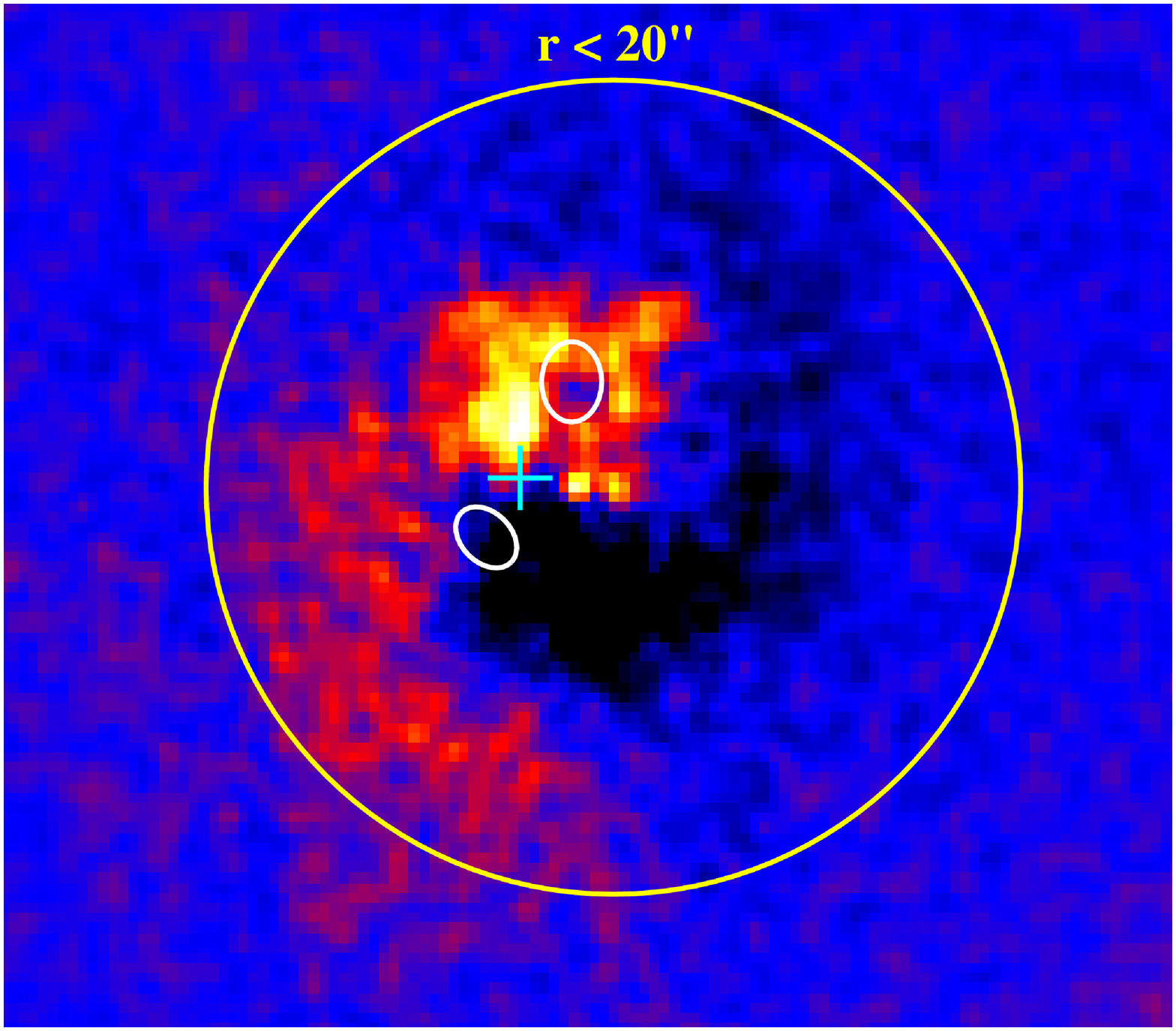}  
\hspace{5mm}
\includegraphics[width=70mm, height=70mm]{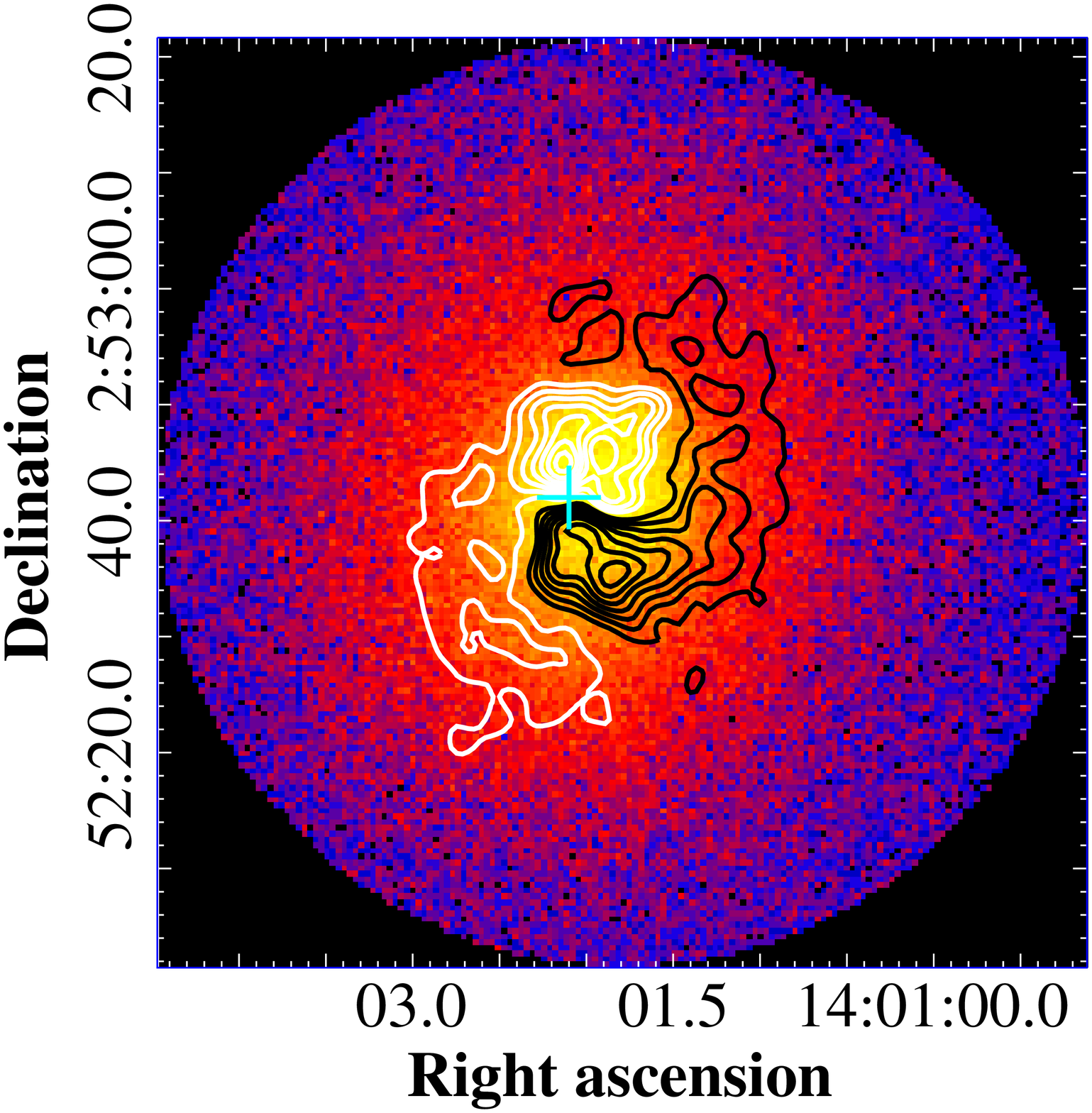}  
\caption{
Spiral patterns in the central region of A1835.
Left panel: X-ray residual image of the central region of A1835, smoothed with a $2\arcsec$ Gaussian.
Pronounced spiral patterns are found. The positive-excess region extends from north to south and the negative-excess region is at the opposite side to the center.
The white circle of $r < 20\arcsec$ is overlaid as a reference.
The cyan cross shows the position of the BCG of A1835.
The two white ellipses show the position and size of X-ray cavities reported by \cite{McNamara06}.
Right panel: same as Figure~\ref{fig:image} but the contours of the positive (in white) and negative (in black) excess regions are overlaid.
White contours show the excess levels of 100\,\%, 90\,\%, $\cdots$, 20\,\%, 10\,\% of the peak value of the positive brightness residuals. 
Black contours are the same as the white contours but for the negative residuals. 
Both the spiral patterns are embedded in the cool core and are not distinguishable in the original X-ray image.
}
\label{fig:spiral}
\end{center}
\end{figure*}

Then, we subtract the obtained model surface brightness from the original X-ray image.
The resultant image is smoothed with a 2\arcsec ~Gaussian and shown in the left panel of Figure~\ref{fig:spiral}.
We find a distinctive positive-excess region, which extends from north to south counterclockwise, as well as a negative-excess region at the opposite side of the cluster center in a nearly symmetric shape, 
i.e.,  the two regions form spiral patterns.
We plot two white ellipses in the left panel of Figure~\ref{fig:spiral}, 
which show the position and size of X-ray cavities reported by \cite{McNamara06}, 
read off from Fig. 9 and Table 2 of \cite{McNamara06}.
The right panel of Figure~\ref{fig:spiral} shows the distribution of the X-ray surface brightness of the spiral patterns as a contour map overlaid on the original X-ray image of Figure~\ref{fig:image} right panel.
We define the size of the spiral pattern as the distance from the
peak position to the farthest tail of the contour that corresponds to 10\,\% of the peak residual brightness shown in Figure~\ref{fig:spiral}.
Thus, the spiral patterns as a whole extend to a circular area with the radius of $\sim 18\arcsec$ (or $\sim 70$\,kpc) in the residual image of A1835.
Moreover, the size of the spiral with a positive-excess is consistent with that of the spiral with a negative-excess.

\begin{figure}[ht] 
\begin{center}
\includegraphics[height=40mm, width=75mm]{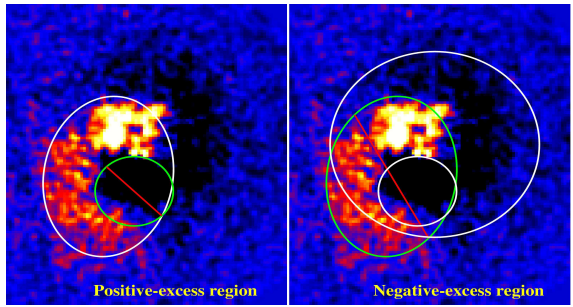}  
\caption{
The regions from which the X-ray spectra of the positive ({\it left}) and negative ({\it right}) excess regions are extracted. 
X-ray spectra are accumulated from inside the region in white line excluding the smaller region in green line with a red diagonal line in each panel.
}
\label{fig:spiralreg}
\end{center}
\end{figure}

\begin{figure*}[ht] 
\begin{center}
\includegraphics[width=72mm, height=54mm]{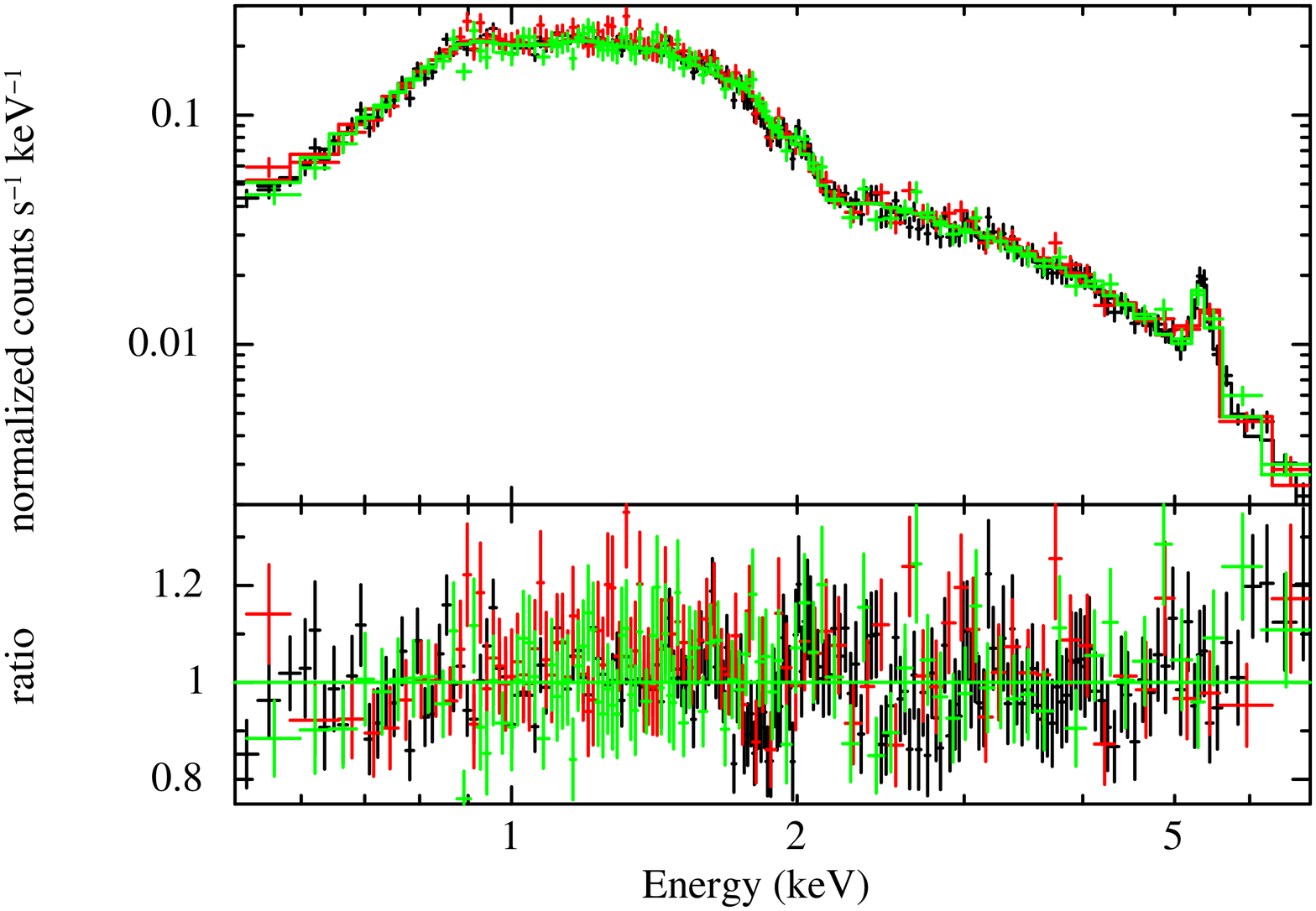}  
\hspace{2mm}
\includegraphics[width=72mm, height=54mm]{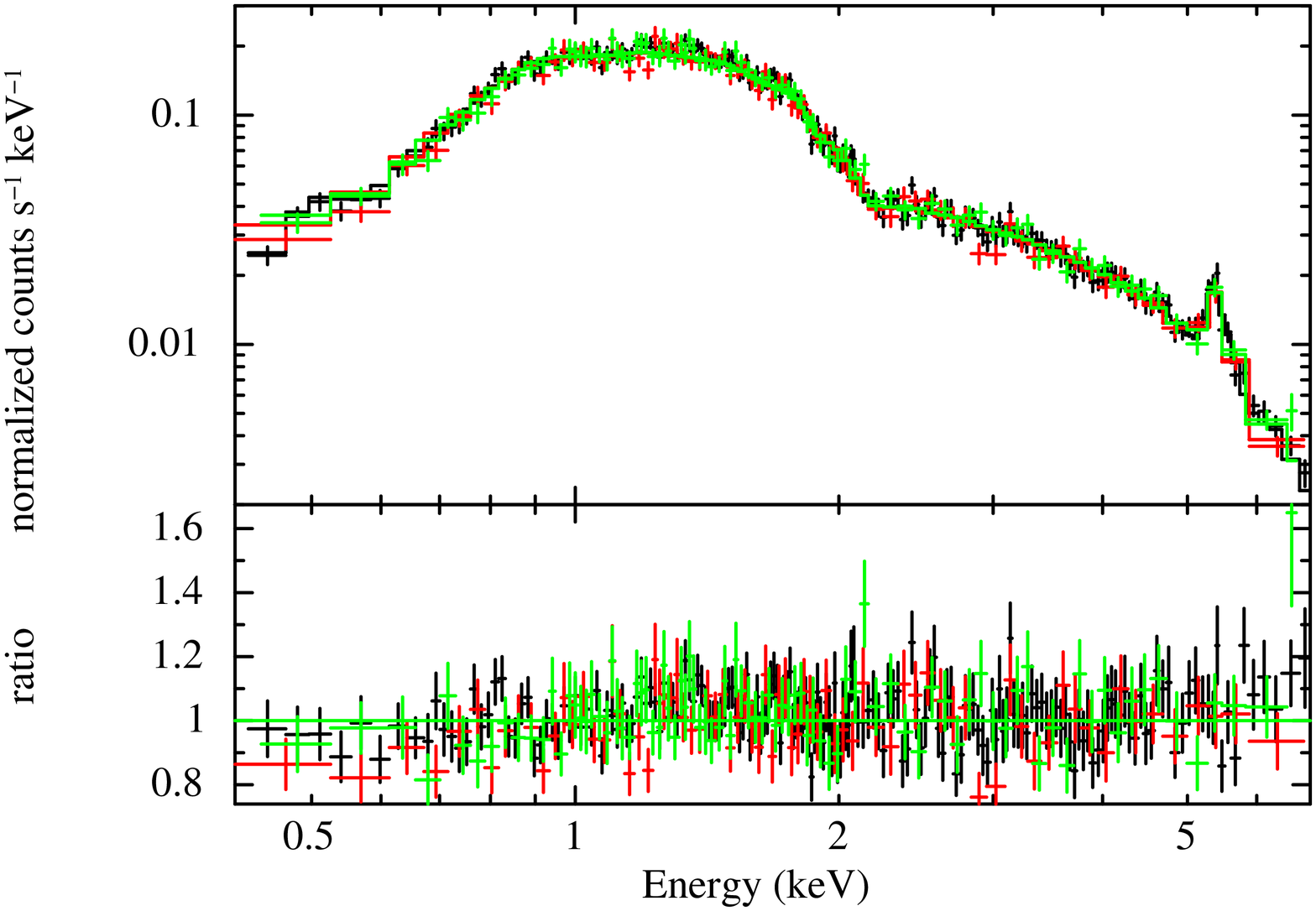}  
\caption{
X-ray spectra extracted from the positive-excess region ({\it left}) and from the negative-excess region ({\it right}) for each ObsID are shown with the best-fit APEC models.
The ratios of the data to the model are plotted in the bottom panels.
The spectra in black, red, and green are extracted from the data of ObsID $= 6880$,  6881, and  7370, respectively.
All of the X-ray spectra are well reproduced with the APEC model.
}
\label{fig:spec}
\end{center}
\end{figure*}

Next, we extract X-ray spectra of the positive- and negative-excess regions of the spiral patterns from the regions as marked in Figure~\ref{fig:spiralreg}.
We fit them using XSPEC version 12.9.0o \citep{Arnaud96}.
In the spectral fitting, the Galactic absorption (\NH) is fixed to $2.04 \times 10^{20}$\,cm$^{-2}$ \citep{Kalberla05} and the ICM emission is assumed to be reproduced by a single temperature model 
of a thin-thermal plasma \citep[APEC;][]{Smith01}. 
Figure~\ref{fig:spec} shows the fitting results of the X-ray spectra of the positive ({\it left}) and the negative ({\it right}) excess regions for each ObsID;
the spectra in black, red, and green are extracted from the data of ObsID $= 6880$, 6881, and 7370, respectively.
The ICM spectra in the regions of the spiral patterns are well described by the single temperature model with $\chi^{2}/{\rm dof}$ of 778/706 and 812/716 for the  positive- and negative-excess regions, respectively.

The ICM temperatures of the positive- and negative- excess regions are found to be  $5.09 ^{+0.12} _{-0.13}$\,keV and $6.52 ^{+0.18} _{-0.15}$\,keV, respectively.
The respective abundances ($Z$) are $0.37 ^{+0.04} _{-0.03}$\,$Z_{\odot}$ and $0.39 ^{+0.04} _{-0.04}$\,$Z_{\odot}$.
The electron density, pressure, and entropy are computed assuming that the ICM is distributed uniformly over the line-of-sight of length $L$.
Assuming a factor of density correction of $n_{\rm e} / n_{\rm H} = 1.17 + 0.02 \times Z$ (with \cite{Anders89}, the primordial Helium abundance $Y_{p} = 0.25$), 
the electron density is estimated to be $n_{\rm e} = 0.0169 \pm 0.0003$\,cm$^{-3}$\,$(L/1\,{\rm Mpc})^{-1/2}$ and $n_{\rm e} = 0.0120 \pm 0.0003$\,cm$^{-3}$\,$(L/1\,{\rm Mpc})^{-1/2}$ for the former and latter, respectively, where $L$ denotes the depth of line-of-sight.
Then, the electron pressure and entropy of the positive-excess region are $0.086 \pm 0.003$\,keV\,cm$^{-3}$\,$(L/1\,{\rm Mpc})^{-1/2}$ and $77.3 ^{+2.1} _{-2.2}$\,keV\,cm$^{2}$\,$(L/1\,{\rm Mpc})^{1/3}$, 
and those of the negative-excess region are $0.078 ^{+0.003} _{-0.002}$\,keV\,cm$^{-3}$\,$(L/1\,{\rm Mpc})^{-1/2}$ and $124.2 ^{+3.8} _{-3.3}$\,keV\,cm$^{2}$\,$(L/1\,{\rm Mpc})^{1/3}$, respectively.
Note that these are not real electron density, pressure, or entropy, as we did not perform any deprojection to estimate a 3-dimensional distribution.
The measured redshifts are consistent among these fitting results and cluster's one.
We find no significant bulk motions in the direction of line-of-sight with the upper limit for the velocity of ICM motion in the redshift difference ($\Delta z$) of 0.002, which corresponds to 600\,km\,s$^{-1}$.
We summarize the best-fit results in Table~\ref{tab:fit}.

\begin{table}[ht]
\begin{center}
\caption{
Best-fit parameters of X-ray spectra in the spiral patterns.
}\label{tab:fit}
\begin{tabular}{ccc}
\hline\hline	
Region									&	 Positive-excess							&	 Negative-excess							\\ \hline
Temperature (keV)				&	$5.09 ^{+0.12} _{-0.13}$		&	$6.52 ^{+0.18} _{-0.15}$		\\
Abundance ($Z_{\odot}$)				&	$0.37 ^{+0.04} _{-0.03}$		&	$0.39 ^{+0.04} _{-0.04}$		\\
Redshift								&	$0.251 ^{+0.003} _{-0.004}$	&	$0.249 ^{+0.004} _{-0.004}$	\\
Density (cm$^{-3}$\,$(L/1\,{\rm Mpc})^{-1/2}$)			&	$0.0169 \pm 0.0003$				&	$0.0120 \pm 0.0003$				\\ 
$\chi^{2}$/dof					&	778/706										&	812/716										\\ \hline
\multicolumn{3}{c}{}	\\	\hline
Pressure (keV cm$^{-3}$\,$(L/1\,{\rm Mpc})^{-1/2}$)	&	$0.086 \pm 0.003$					&	$0.078 ^{+0.003} _{-0.002}$	\\
Entropy (keV cm$^{2}$\,$(L/1\,{\rm Mpc})^{1/3}$)	&	$77.3 ^{+2.1} _{-2.2}$			&	$124.2 ^{+3.8} _{-3.3}$			\\
\hline
\end{tabular}
\end{center}
\end{table}

\begin{figure}[t] 
\begin{center}
\includegraphics[width=66mm, height=200mm]{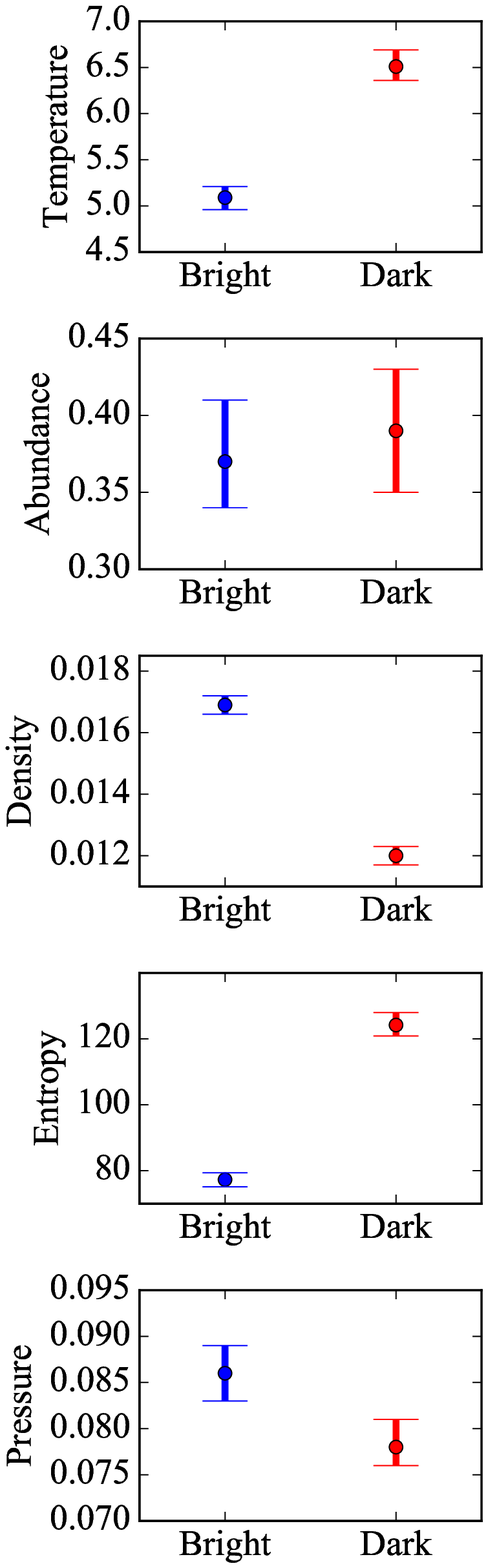} 
\caption{
Best-fit parameters of spectral analyses in the positive- (left side in each panel) and negative- (right side) excess regions.
The panels from the top to bottom show the temperature in units of keV, abundance in  solar, density in cm$^{-3}$\,$(L/1\,{\rm Mpc})^{-1/2}$, entropy in keV cm$^{2}$\,$(L/1\,{\rm Mpc})^{1/3}$,
and pressure in  keV cm$^{-3}$\,$(L/1\,{\rm Mpc})^{-1/2}$, respectively.
}
\label{fig:fit}
\end{center}
\end{figure}

Figure~\ref{fig:fit} shows the properties of the ICM in the regions of the spiral patterns.
Although the temperature, density, and entropy differ between the two regions,
the pressures of the ICM between them are very similar.  
It suggests that they are in pressure equilibrium or close.

\subsection{Spectral analyses of surroundings}
\label{sec:sur}

We have measured the ICM properties in the regions of the spiral patterns in the previous subsection.
In this subsection, we perform spatially-resolved detailed X-ray spectroscopy of the entire cluster-center region ($r < 40\arcsec$ or $r < 160$\,kpc) including the spiral patterns.

We first apply the contour-binning algorithm (hereafter ContBin) presented by \cite{Sanders06} to determine how to divide the region into many subregions to extract X-ray spectra from. 
This algorithm yields a set of sub-regions so that each region has roughly the same level of signal-to-noise ratio (S/N).
We divide the whole region (right panel of Figure~\ref{fig:image}) into 92 sub-regions
with more than $\sim 2000$\,counts in each region after subtracting the background.
The S/N, where the Poisson noise N includes the background, of each region is within 42.2 -- 57.6 in the 0.4 -- 7.0\,keV band.
Then, we make X-ray spectrum in each region.
By fitting each spectrum with the APEC model, we obtain the spatial distribution of the ICM temperature, abundance, and density. 
The pressure and entropy in each region are then calculated.

\begin{figure*}[ht] 
\begin{center}
\includegraphics[width=74mm, height=56mm]{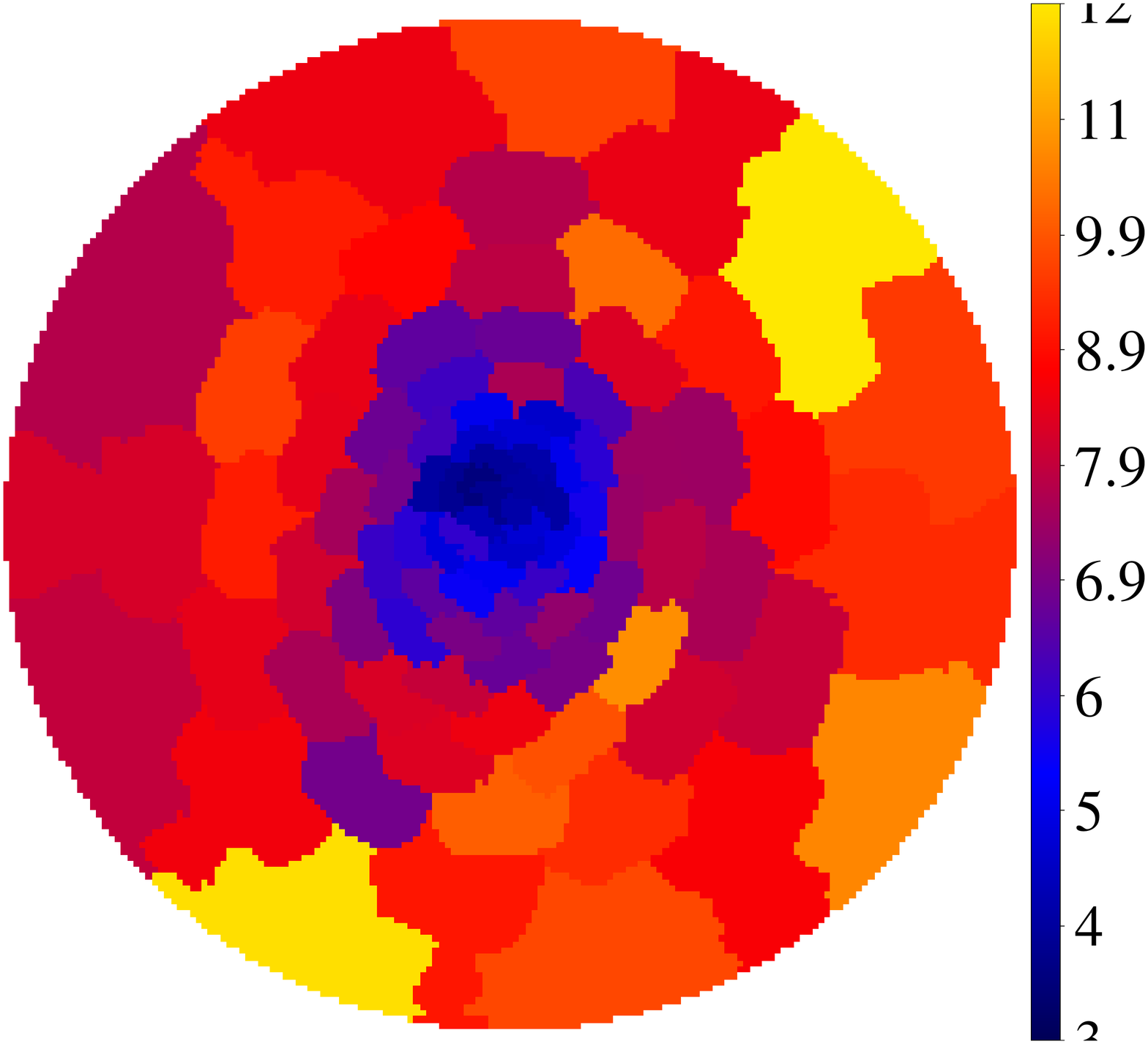}  
\hspace{2mm}
\includegraphics[width=74mm, height=56mm]{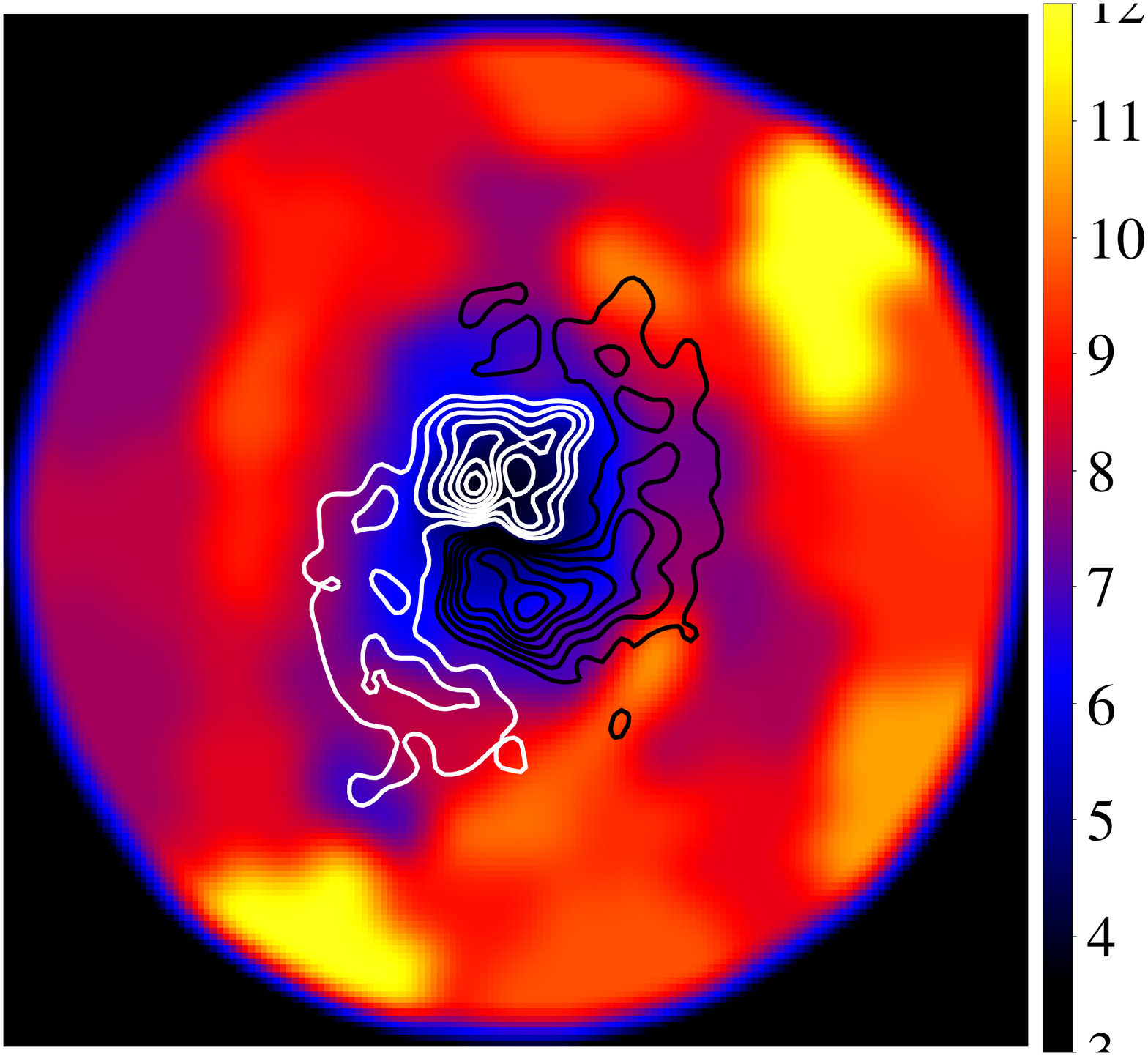}  
\caption{
Temperature maps obtained by X-ray spectral analyses with the ContBin-defined regions in units of keV.
The left panel shows the raw temperature map.
The statistical error of the inner/outer regions in the map are 7\,\%/15\,\%, respectively.
The right panel shows the smoothed image of the left panel with $7\arcsec$ Gaussian, together with the contours overlaid of the spiral patterns as in the right panel of Figure~\ref{fig:spiral}.
}
\label{fig:kT}
\end{center}
\end{figure*}

\begin{figure*}[ht] 
\begin{center}
\includegraphics[width=74mm, height=56mm]{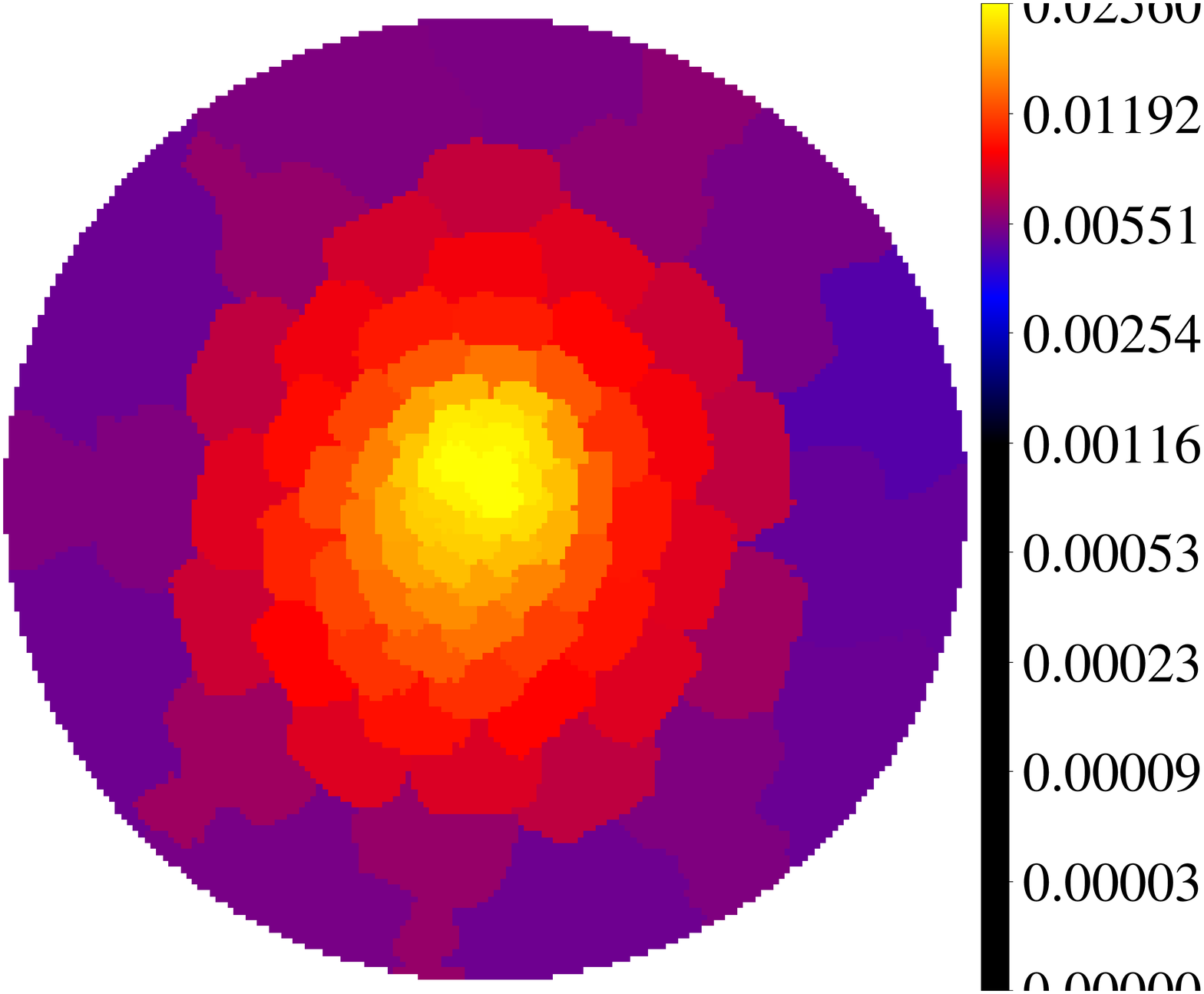}  
\hspace{2mm}
\includegraphics[width=74mm, height=56mm]{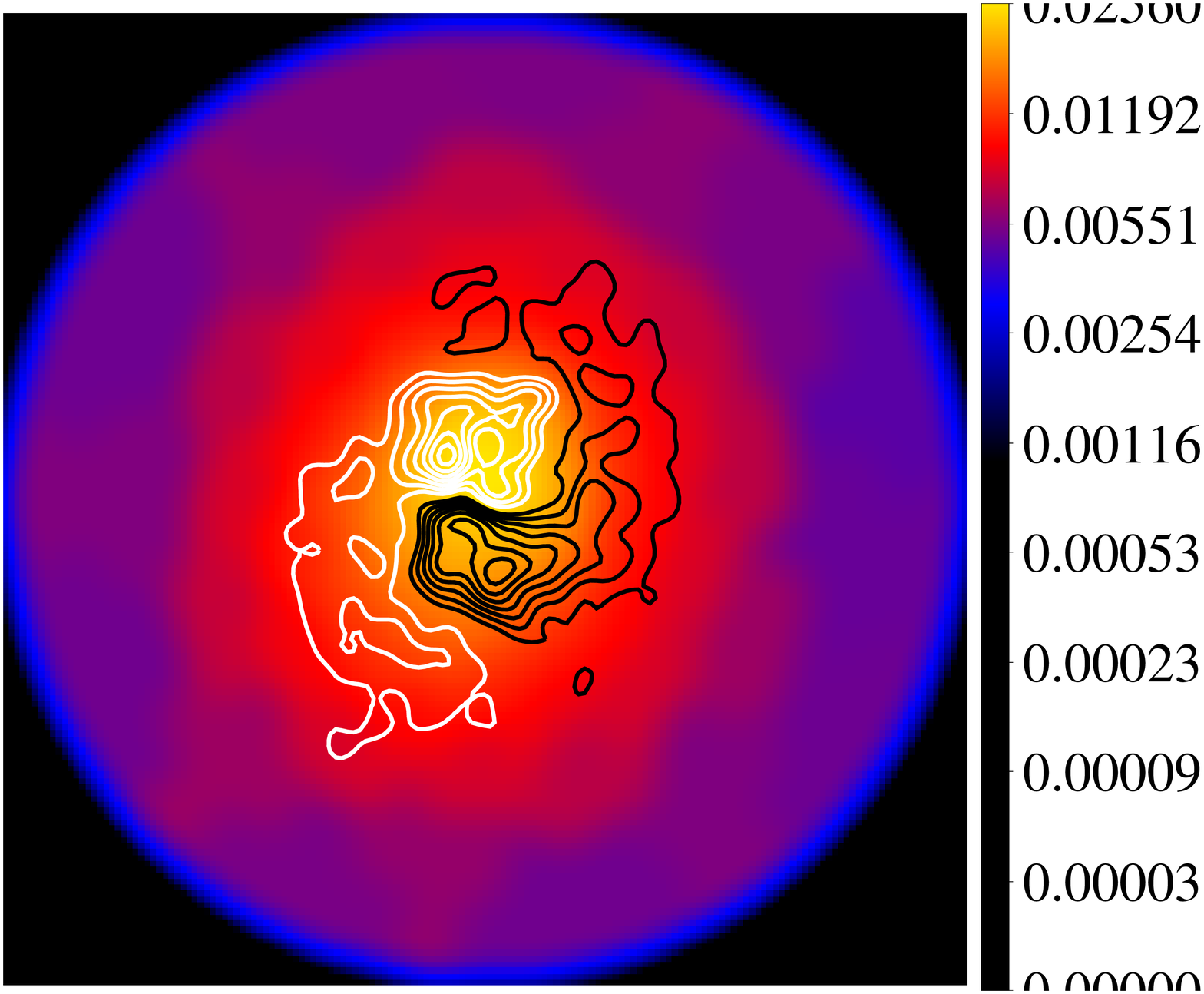}  
\caption{
Same as in Figure~\ref{fig:kT} but for the density maps in units of cm$^{-3} ~(L/1\,{\rm Mpc})^{-1/2}$.
The statistical error is 5\,\%\ for both the inner and outer regions.
The right panel shows the same image as the left panel but that is smoothed with $7\arcsec$ Gaussian and the contours of the spiral patterns shown in the right panel of Figure~\ref{fig:spiral} are overlaid.
}
\label{fig:ne}
\end{center}
\end{figure*}

\begin{figure*}[ht] 
\begin{center}
\includegraphics[width=74mm, height=56mm]{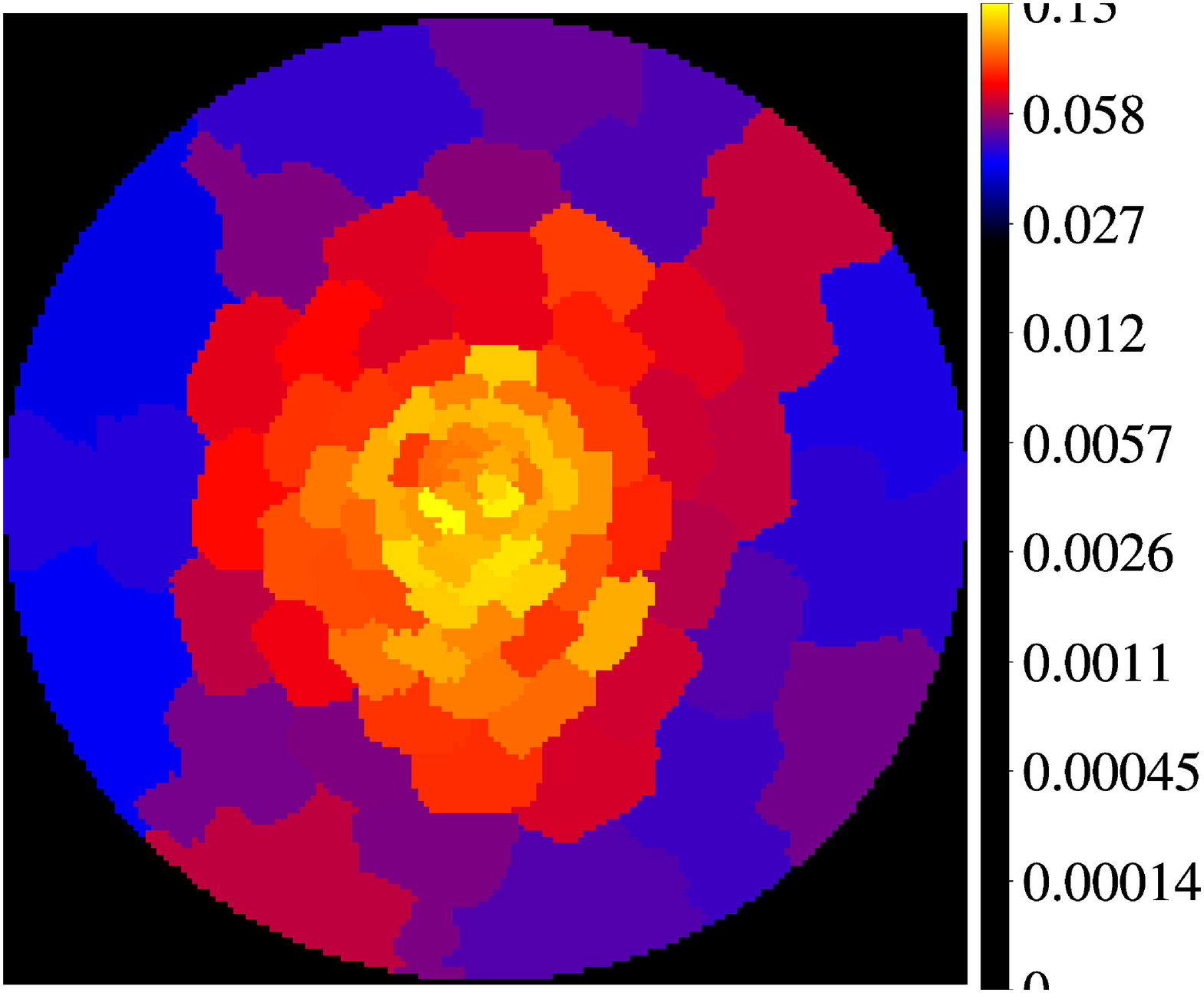}  
\hspace{2mm}
\includegraphics[width=74mm, height=56mm]{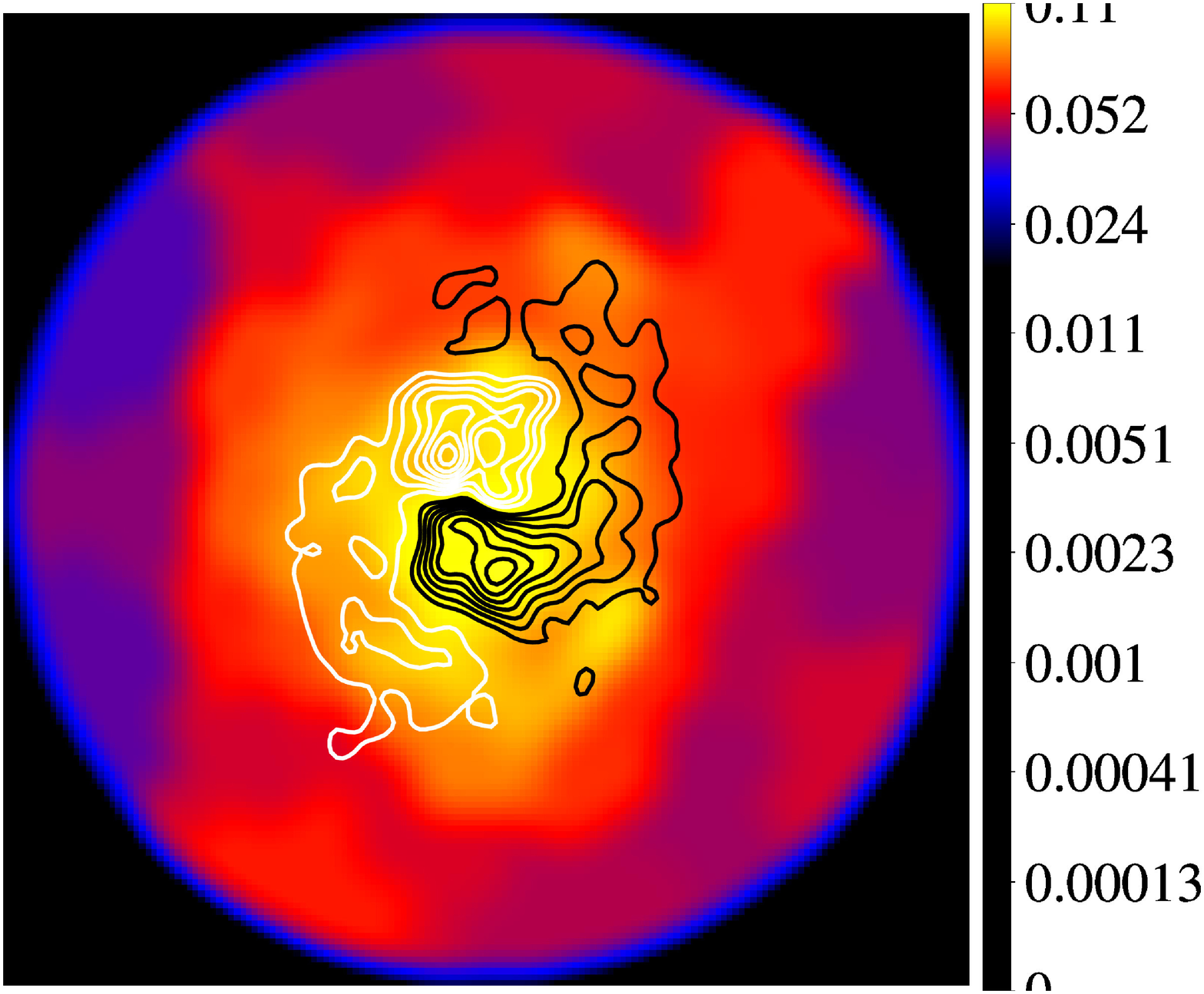}  
\caption{
Same as in Figure~\ref{fig:kT} but for the pressure maps in units of keV\,cm$^{-3} ~(L/1\,{\rm Mpc})^{-1/2}$, which is calculated by $P = kT \times n_{\rm e}$.
The statistical error of inner region in the map is 9\,\% and that in outer region is 16\,\%.
}
\label{fig:pressure}
\end{center}
\end{figure*}

\begin{figure*}[ht] 
\begin{center}
\includegraphics[width=74mm, height=56mm]{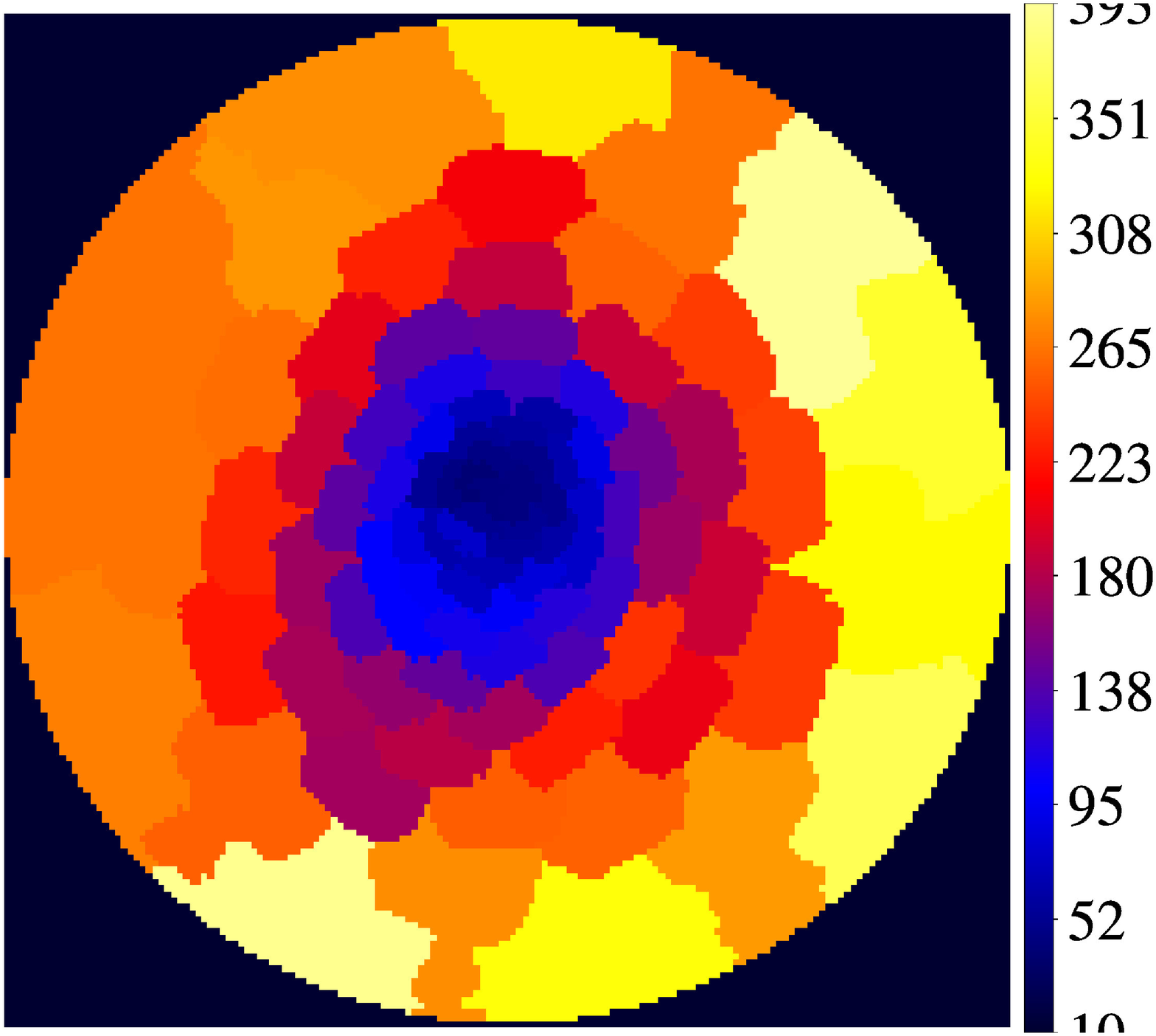}  
\hspace{2mm}
\includegraphics[width=74mm, height=56mm]{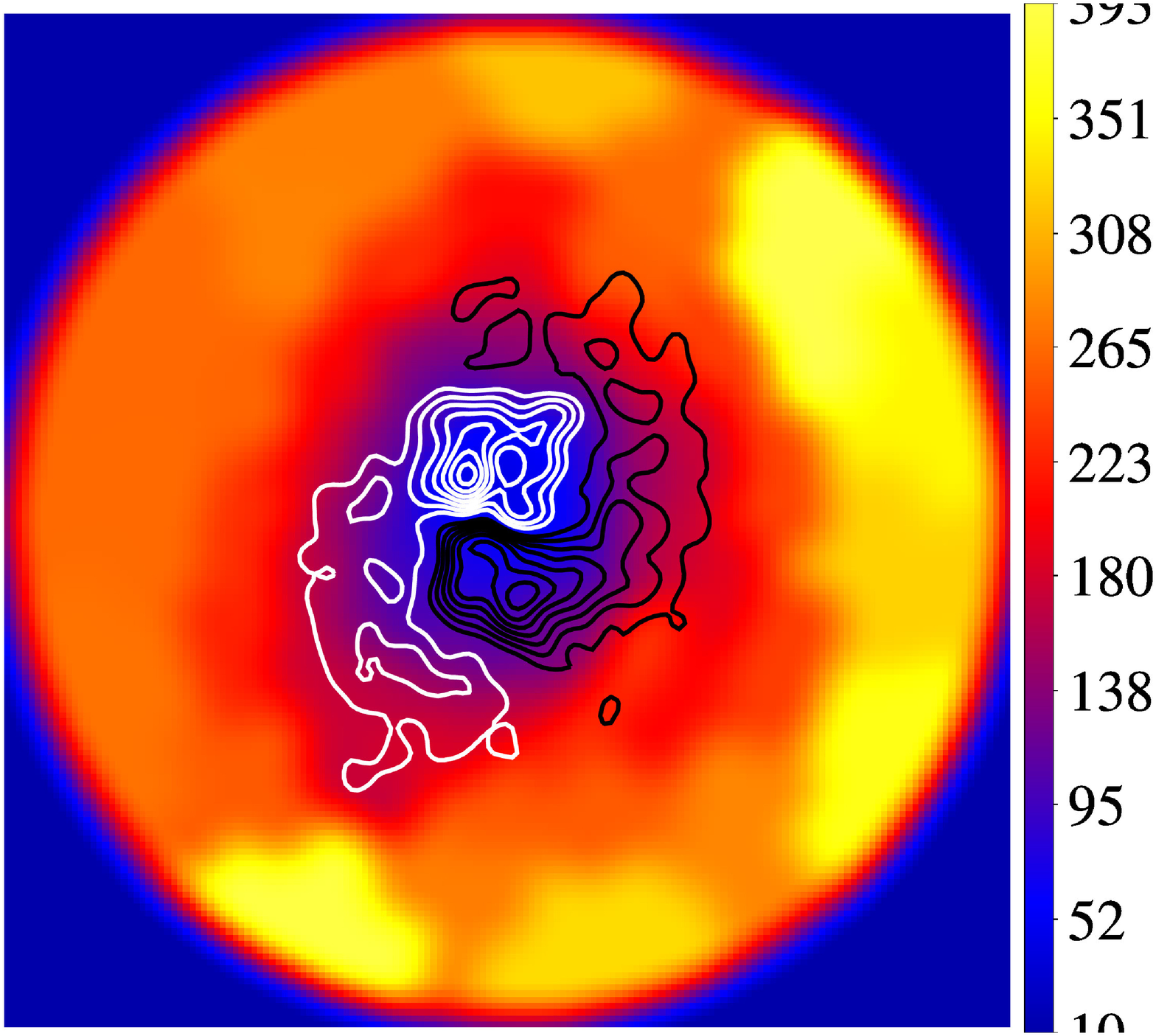}  
\caption{
Same as in Figure~\ref{fig:kT} but for the entropy maps in units of keV\,cm$^{2} ~(L/1\,{\rm Mpc})^{1/3}$, which is calculated by $K = kT \times n_{\rm e}^{-2/3}$.
The statistical error of inner region in the map is 8\,\% and that in outer region is 15\,\%.
}
\label{fig:entropy}
\end{center}
\end{figure*}

We show the ICM temperature, density, pressure, and entropy maps in Figure~\ref{fig:kT}, \ref{fig:ne}, \ref{fig:pressure}, and \ref{fig:entropy}, respectively.
The statistical errors in the inner/outer regions in the temperature map are 7\,\%/15\,\%, respectively.
Those for the density, pressure, and entropy are 5\,\%/5\,\%, 9\,\%/16\,\%, and 8\,\%/15\,\%, respectively, for the inner/outer regions.

The temperature map (Figure~\ref{fig:kT}) shows that the ICM in the center is the coolest at 3.5\,keV in the center and becomes gradually hotter at a larger radius.
In the outermost region ($35\arcsec < r < 40\arcsec$), the lowest and highest temperatures are 7.7\,keV and 11.9\,keV, respectively.
We find that the lower temperature region is elongated toward the southeast direction, which roughly traces the positive-excess region of the spiral patterns (right panel of Figure~\ref{fig:kT}).
Most of the negative-excess region of the spiral patterns has a relatively higher temperature than that of the positive-excess region.
This is consistent with the picture provided by the spectral analyses of the spirals.
It is noteworthy that the spiral patterns appear in the relatively lower temperature regions.

The abundance map is effectively flat and its global patterns is consistent with that reported by \cite{Kirkpatrick15}.

The density map (Figure~\ref{fig:ne}) traces well the shape of X-ray surface brightness of A1835, i.e., it is elongated to the southeast with the peak at the cluster center.

The pressure map (Figure~\ref{fig:pressure}) shows that the ICM pressure smoothly decreases toward the outer region with no significant discontinuities in the central region.
In the region of the spiral patterns, the pressure profile shows no discontinuity, which is consistent with that obtained by the X-ray spectral analyses of spirals.
These results imply that the ICM is in pressure equilibrium or close at least in the region of the spiral patterns.

The entropy map (Figure~\ref{fig:entropy}) shows the entropy of the ICM is lowest at the center.
It also shows that the spatial distribution of the lower entropy ICM is elongated to the southeast, being similar to that of the positive-excess region of the spiral patterns.
Note that this result is consistent with the direct spectral analyses of the spiral patterns presented in the previous subsection, where
the spectrum of the negative-excess region, which is located in the western side, has shown a higher entropy than that of the positive-excess region in the eastern side.

\section{Discussion}

We have found the embedded spiral patterns in the residual image of X-ray surface brightness of the core of A1835.
We then derived the spectral properties of the ICM in them and in their surroundings.
In this section, we discuss the results and their implications.

In previous works, by using the data of 33 clusters, \cite{Hofmann16} studied the large-scale (typically over hundreds kpc) perturbations of X-ray surface brightness. 
A1835 is one of their samples and its unsharp-masked count image is shown in Fig. E.1 in \cite{Hofmann16}.
Since they did not focus on the central region (less than 100\,kpc), the spatial resolution of their unsharp-masked count image is low.
The feature such as the spirals we found is not apparent in their results.

\subsection{Properties of spiral patterns}
\label{sec:property}

We have estimated the size of the spiral patterns to be 70\,kpc in radius (Figure~\ref{fig:spiral}).
This size is a factor of  2 -- 4 smaller than those of other known clusters, e.g., about 200\,kpc for A496 \citep{Ghizzardi14}, 300\,kpc for A2029 \citep{Clarke04}, and 240\,kpc for A2052 \citep{Blanton11}.
To compare the size of the spiral patterns in more objective manner, 
we re-analyze the 650\,ksec data of A2052 observed with {{\it Chandra},
which are the same as those used in \cite{Blanton11}.
Then, we measure the size of the spiral patterns in A2052 using the same method as described in Section~\ref{sec:search} to be 170\,kpc.
For A496, from Fig.~13 of \cite{Ghizzardi14},
we measure the size of spiral to be 150\,kpc by using the same method as in Section~\ref{sec:search}.
These results further confirm that the size of the spiral patterns of A1835 is smaller.
We will perform more systematic investigations of the spiral patterns for a larger sample of clusters in our future publication.  

We also measure the core radius of A1835 to be $31 \pm 0.4$\,kpc using our a single $\beta$ model. 
It is within the range of the core radii of the other known clusters with spiral patterns; the core radii of the above-mentioned three clusters of A496, A2029, and A2052 are 15\,kpc \citep{Lagana08}, 38\,kpc \citep{Lewis03}, and 18\,kpc \citep{Blanton11}, respectively. 
We find no obvious correlation between size of spirals and that of core radius.
In fact, the size of spiral patterns of A1835 is much closer to the core radius than the other clusters; they extend from well inside the cool core out to the hotter surrounding ICM.  
This may indicate that the stirring motion producing the spirals transports a fraction of the cool gas out to larger radii and the hot gas into smaller radii.

We have also analyzed the X-ray spectra extracted from the positive- and negative-excess regions of the spiral patterns (Figure~\ref{fig:spiralreg}) and have measured the spatial distribution of the ICM parameters (Table~\ref{tab:fit} and Figure~\ref{fig:fit}).
We find that the difference and similarity of the ICM parameters between the two regions are qualitatively in good agreement with those of other known clusters \citep{Clarke04, Blanton11, Paterno-Mahler13, Ghizzardi14},
i.e., the temperature and entropy of the ICM in the positive-excess region are lower than those in the negative-excess region, and the ICM density in the  former is higher than that in the latter.
Whereas the pressure has no difference between the two regions.
The spectral results of their surrounding area has also confirmed the uniform pressure  across the region of the spiral patterns.
Therefore, the ICM in the region of the spiral patterns is in pressure equilibrium or close.
For the temperature map, \cite{Hofmann16} studied the large-scale distribution in A1835. 
They reported that the temperature of the outer region exceeds 10\,keV and its distribution is elongated toward the northwest and southeast.

For the abundance, we have found no difference between the positive- and negative-excess regions.
By combining the results of spectral analyses of the spirals and of their surroundings, we find almost uniform distribution of metals across the region with a radius of at least $r < 40\arcsec$ in A1835.
This is different from that reported in the previous studies of the spiral patterns.
For example, in the case of A496, the abundance of the positive-excess region is larger by a factor of $\sim$1.5 than that of the negative-excess region \citep{Ghizzardi14}.

We consider two possible scenarios for the origin of the spiral patterns: an energetic past AGN activity, including jets, and a gas sloshing induced by an off-axis minor merger \citep[e.g.,][]{Markevitch07}.
\cite{McNamara06} pointed out the presence of X-ray cavities in the core and suggested that the energetic AGN outburst had occurred 40\,Myr ago.
They also suggested that the past jet activity had suppressed a cooling of most of the ICM.
Such an energetic activity can influence the surroundings close to the center as in the case of the Perseus cluster \citep{Fabian11}.
However, \cite{McNamara06} estimated the size of the X-ray cavity to be $\sim 20$\,kpc, which is significantly smaller than that of the spiral patterns that we found (left panel of Figure~\ref{fig:spiral}).
It is therefore more likely that the origin is the same as in the other clusters, i.e., the spiral patterns are created by the gas sloshing.
If so, since it is considered that an off-axis minor merger induces the gas sloshing \citep[e.g.,][]{Markevitch07}.
The presence of the spiral patterns is a piece of the evidence of a past minor merger in A1835.
We should note that a sub-structure exists and must be disturbing the gas in the cool core in A1835, even though the X-ray surface brightness indicates that it is relaxed.

If the origin of the spiral patterns is gas sloshing induced by a minor merger, 
the relatively small size of the spiral patterns may reflect a time evolution of spirals.
According to the numerical simulations \citep[e.g.,][]{ZuHone10}, the size of the spirals depends on the length of the period after a minor merger began.
In the early stage of a minor merger, a volume over which the gas is disturbed is relatively small and the effect of disturbance is limited in the core of the cluster.
However, the conditions of initial merger parameters such as an impact parameter also affect the form and size of spirals.

For the pressure profile in the cluster center, observations of the Sunyaev-Zel'dovich (SZ) effect would enable us to directly measure the pressure of the ICM.
Although the observations of the SZ effect for A1835 were performed by \cite{Korngut11}, the signal in the central part ($r < 18\arcsec$) was strongly contaminated by the AGN emission.
The size of the contaminated region is consistent with that of the spiral patterns.
Now, observations with ALMA, which can achieve a high angular resolution of $5\arcsec$ and high sensitivity for the SZ effect \citep{Kitayama16}, will likely enable us to make an accurate measurement of the pressure of the ICM in the region of the spiral patterns.

\subsection{Spiral patterns and mergers}

Following the simulations of A1835 \citep[e.g.,][]{Ascasibar06, ZuHone10, Roediger11}, we consider an orbital plane of the first passage of the minor merger.
They argue that the low-temperature region (i.e., the positive-excess region) appears at the position opposite to the direction of the trajectory.
Since the positive-excess region of A1835 extends from the north to south counterclockwise, we conjecture that the infalling subcluster traveled from north to southwest.

As discussed above (Section~\ref{sec:property}), the comparatively small size of the spiral patterns of A1835 may be an indication that the length of the period after the beginning of the minor merger is relatively short.
On the other hand, \cite{Smith08} suggested that growth in mass of A1835 in the last 2\,Gyr is smaller than 10\,\%, based on the observations of the mass fraction associated with the sub-structure.
The simulations demonstrated that a subcluster with a mass ratio of 5\,\% can slosh the gas  in the gravitational potential well of the main cluster after the first contact and can then create a temperature structure of spiral patterns \citep[e.g.,][]{ZuHone10}.
For A1835, the mass ratio of 5\,\% corresponds to $\sim 5 \times 10^{13}$\,\MO ~for $M_{200}$ of A1835 \citep{Okabe10}, which is within the range of the mass of groups of galaxies. 
In that case, it satisfies the conditions proposed by \cite{Smith08}.
\cite{Bonamente13} and \cite{Ichikawa13} suggested that the ICM in the outskirts is out of hydrostatic equilibrium.
Recent minor merger that we have guessed A1835 has experienced may have induced it.

The velocity difference of the ICM between the positive- and negative-excess regions is smaller than 600\,km\,s$^{-1}$.
We have found no bulk motions along the line-of-sight.
\cite{Sanders10} reported that the turbulence velocity at the center is smaller than 274\,km\,s$^{-1}$.
Since the ICM does not slosh fast, these two results support the hypothesis that the orbit of the minor merger was in the plane of the sky.
To make an accurate measurement of the velocity of the ICM is beyond the capability of currently available X-ray observatories, but will be possible with future X-ray micro-calorimeter missions.

\cite{Govoni09} reported the presence of a radio mini-halo in A1835.
The radio emission is concentrated in the center and shows no obvious asymmetry.
We do not find direct correspondence between the morphology of the spiral pattern in the X-ray residual brightness and that of the radio mini-halo. 



\section{Summary}

We find the embedded spiral patterns in the cool core in the residual image of X-ray surface brightness after the averaged profile is subtracted in the massive clusters of galaxies A1835 using the data of {\it Chandra X-ray Observatory}.
We measure the size of the spiral patterns to be 70\,kpc, which is a factor of  2 -- 4 smaller than that of other clusters known to show the similar features.
The spiral patterns consist of two arms, which appear as the positive- and negative-excess in the residual image. 
We analyze the X-ray spectra extracted from those two regions separately.
The ICM temperatures in the positive- and negative-excess regions are found to be $5.09 ^{+0.12} _{-0.13}$\,keV and $6.52 ^{+0.18} _{-0.15}$\,keV, respectively.
No significant difference is found in the ICM abundance in the two regions, nor in the pressure, which indicates that the ICM is in pressure equilibrium or close.
We measure the velocity difference of the ICM between the positive- and negative-excess region along line-of-sight to be smaller than 600\,km\,s$^{-1}$.

Applying the algorithm of ContBin, we define the sub-regions so that any of them has S/N $> 42$ in the 0.4 -- 7.0\,keV band (i.e., $\sim 2000$\,photons in each region) in the central region of $r < 40\arcsec$ in A1835 after the background is subtracted.
Compiling the X-ray spectra from the 92 sub-regions, we make the temperature, density, pressure, and entropy maps (Figures~\ref{fig:kT} -- \ref{fig:entropy}). 
The radial profiles of temperature, density, and entropy imply that A1835 is a typical cool-core cluster.
The spiral patterns extend from the cool core out to the hotter surrounding ICM. 
This may indicate that the stirring motion producing the spirals transports a fraction of the cool gas out to larger radii and the hot gas into smaller radii.
The spatial distribution of metals seems to be uniform at least in the regions.
The pressure distribution shows a gradual decrease toward the outskirts and implies that the ICM in A1835 is in pressure equilibrium at least in the central region, even though the spiral patterns exist in the core, which suggests that the core is disturbed.

Two scenarios are considered as the mechanism of disturbance of the cool core.
One is an energetic past AGN activity including jets and the other is gas sloshing induced by an off-axis minor merger.
Although we can not determine conclusively which is more plausible because of the insufficient X-ray data, the ICM properties in the region of the spiral patterns are similar to those in other clusters that show embedded spiral patterns in their X-ray surface brightness.
In either case, the spiral patterns indicate that a sub-structure exists in the cool core even if the X-ray surface brightness appears to be relaxed.
The velocity of disturbance in the line-of-sight is lower than 600\,km\,s$^{-1}$ for A1835, which is consistent with that of the Perseus cluster, as directly measured by {\it Hitomi} \citep{Hitomi16}.

\acknowledgments
We are grateful to the anonymous referee for helpful suggestions and comments.
We also thank Masaaki Sakano for careful reading of the manuscript and useful comments, and Yuto Ichinohe for helpful discussions.
The scientific results of this paper are based in part on the data obtained from the Chandra Data Archive: ObsIDs 6880, 6881, and 7370.
This research is supported by Japan Society for the Promotion of Science (JSPS) KAKENHI Grant Number 15K17610 (SU), 25400236 (TK), and 24105007 (TD).

\vspace{5mm}


\bibliographystyle{apj}
\bibliography{00_BibTeX_library}



\end{document}